# Radial bound states in the continuum for polarization-invariant nanophotonics

Lucca Kühner[1,2], Luca Sortino[1,2], Rodrigo Berté[1,2,3], Juan Wang[1,2], Haoran Ren[4],

Stefan A. Maier[4,5,1,2], Yuri Kivshar[6] and Andreas Tittl[1,2,*]

[1]Chair in Hybrid Nanosystems, Nanoinstitute Munich, Faculty of Physics, Ludwig-Maximilians-Universität München, Königinstrasse 10, 80539 Munich, Germany

[2]Center for NanoScience (CeNS), Faculty of Physics, Ludwig-Maximilians-Universität München, Schellingstrasse 4, 80799 Munich, Germany

[3]Instituto de Física, Universidade Federal de Goiás, 74001-970 Goiânia-GO, Brazil

[4]School of Physics and Astronomy, Monash University, Clayton, Victoria 3800, Australia

[5]The Blackett Laboratory, Department of Physics, Imperial College London, London, SW7 2AZ, United Kingdom

[6]Nonlinear Physics Centre, Research School of Physics Australian National University, Canberra ACT 2601, Australia

*e-mail: andreas.tittl@physik.uni-muenchen.de

## Abstract

**All-dielectric nanophotonics underpinned by the physics of bound states in the continuum (BICs) have demonstrated breakthrough applications in nanoscale light manipulation, frequency conversion and optical sensing. Leading BIC implementations range from isolated nanoantennas with localized electromagnetic fields to symmetry-protected metasurfaces with controllable resonance quality (Q) factors. However, they either require structured light illumination with complex beam-shaping optics or large, fabrication-intense arrays of polarization-sensitive unit cells, hindering tailored nanophotonic applications and on-chip integration. Here, we introduce radial quasi bound states in the continuum (radial BICs) as a new class of radially distributed electromagnetic modes controlled by structural asymmetry in a ring of dielectric rod pair resonators. The radial BIC platform provides polarization-invariant and tunable high-Q resonances with strongly enhanced near fields in an ultracompact footprint as low as 2 μm². We demonstrate radial BIC realizations in the visible for sensitive biomolecular detection and enhanced second-harmonic generation from monolayers of transition metal dichalcogenides, opening new perspectives for**

**compact, spectrally selective, and polarization-invariant metadevices for multi-functional light-matter coupling, multiplexed sensing, and high-density on-chip photonics.**

**Introduction**

The rise of optical metasurfaces has launched a variety of breakthrough applications ranging from negative refraction[1] and ultrathin optical elements[2] to photonic computation[3]. Similarly, bound states in the continuum[4] (BICs) have been shown to underpin many fundamental oscillatory phenomena[5] and can be employed for tailoring the lifetimes of resonant wave systems[6]. Initially discovered in quantum physics[7,8] and also found in acoustics[9,10] and ocean science[11], BICs have emerged as an intriguing concept in optics[12–14]. In all-dielectric subwavelength structures, radiative losses of BIC-based systems can be precisely controlled by tailoring interferences of the constituent resonant modes, producing sharp resonances with extremely high values of the quality (Q) factors. The BIC concept has therefore been employed for a variety of applications driven by such spectrally selective nanosystems, including multiplexed biospectroscopy[15–18], high-harmonic generation[19,20], and subwavelength lasing[21–23].

Two of the principal BIC implementations in nanophotonics are supercavity modes in isolated structures such as individual nanoantennas [24,25] and resonances in symmetry-protected metasurfaces[26]. Although single BIC structures offer a minimal footprint, careful electromagnetic engineering is required to satisfy the demanding mode interference conditions of the hybridization of the Mie and Fabry-Pérot modes[24], restricting design flexibility and limiting the range of achievable Q factors. Based on these design constraints, such interference-type modes can only be excited with structured light illumination, further increasing experimental complexity. Moreover, the resonantly enhanced electric near fields are predominantly confined inside the structures, which is advantageous for material-intrinsic processes such as higher harmonic generation, but severely limits surface-driven light-matter interactions and sensing applications.

In contrast, metasurfaces based on all-dielectric resonators with broken in-plane inversion symmetry (Fig. 1a) offer both high Q factors and strong near fields that extend considerably outside of the resonators, enabling cutting-edge applications in sensing[16,18] and enhanced light-matter interaction[20,27,28]. Significantly, symmetry-broken quasi-BICs (qBICs) provide straightforward tunability of the resonance position and Q factor via the geometrical scaling factor of the unit cell[16] and the degree of asymmetry for increased design versatility. However, they require specific excitation polarizations and the lateral on-chip footprints of metasurfaces supporting qBIC resonances in the visible spectrum usually exceed 100 $\mu m^2$ [18,19], holding back their potential for device miniaturization, on-chip multiplexed sensors, and interactions with hybrid nanophotonic systems such as micron-sized individual layers of two-dimensional (2D) materials. Although their spatial extent can be reduced by transforming the system into a one-dimensional chain (Fig. 1b), this comes at a significant cost in terms of Q factor[29] (see Supplementary Note 1).

Here, we introduce the concept of radial bound states in the continuum (radial BICs) as a multi-application platform for sustaining polarization-invariant high-Q resonances with high surface sensitivity in a compact footprint. Notably, ring structures composed of as few as 12 unit cells can efficiently couple incident light into a pronounced radial quasi-BIC mode while providing flexible resonance tuning and strong surface-confined near fields. In our experimental realization, we achieve high Q factors exceeding 500 in the red part of the visible spectrum and demonstrate the viability of our approach as a versatile BIC-based platform for biomolecular sensing and enhanced second harmonic generation in atomically thin 2D semiconductor molybdenum diselenide ($MoSe_2$) monolayers, all in a spatial footprint as small as 2 $\mu m^2$.

**Results**

**Design and implementation of the radial BIC concept**

The symmetry-protected and radially distributed electromagnetic BIC states are accessed through a carefully designed ring structure incorporating symmetry-broken double rod unit cells, where individual resonators are rotated to satisfy radial alignment (Fig. 1c). At resonance, adjacent rods support opposing electric dipoles, which leads to a suppression of the total dipole moment and consequently of the radiative losses in the system. In contrast to previous 2D or 1D resonator arrangements, our approach leverages a semi-infinite ring geometry[30,31], providing a polarization-invariant optical response and avoiding edge effects, which allows a higher number of resonator elements to participate in the qBIC state for increased resonance quality[32]. Owing to the unique circular arrangement, the coupling between elements over the entire ring is improved as well, resulting in much higher Q factors compared to linear resonator chains (Fig. S4). As a result, the radial BIC platform provides the highest quality factor to footprint ratio amongst comparable BIC-based nanophotonic approaches[18–20,32] (Fig. 1). Furthermore, even though neighboring resonators are set at a small angle with respect to each other, the radial arrangement of rods with identical lengths produces a non-radiating BIC state with negligible coupling to the far field (Fig. 2a).

One of the key parameters to control the optical characteristics of the radial BIC mode is the asymmetry parameter $\Delta L$, i.e., the length difference between the two constituent rods of the unit cell. In the symmetric case ($\Delta L = 0$), we observe a BIC-like state, which is decoupled from the radiation continuum and exhibits negligible electromagnetic near-field enhancement (Fig. 2a). For non-zero asymmetry ($\Delta L > 0$), the BIC transforms into a radial quasi-BIC (radial BIC) mode, which can be accessed from the far field, demonstrating strong field enhancements and the characteristic field pattern of oppositely oriented electric dipole moments observed in asymmetric metasurfaces[26] (Fig. 2b). Due to the engineered inter-resonator coupling within the ring, linearly polarized light efficiently couples to the full radial BIC mode extended along the ring and excites spectrally distinct resonances with high Q factors (Fig. S1). Our numerical calculations confirm the presence of highly enhanced near fields outside the resonator volume and highlight how the field magnitude can be tailored via the structural asymmetry (Fig. 2b). We engineer the radial BIC system to exhibit high-Q resonances at around 770 nm and find that the corresponding numerically simulated Q factors exceed 8000 towards the smallest asymmetries (Fig. S2). Furthermore, we observe a clear correlation between the electric near fields (Fig. 2b, for field intensities see Fig. S6) and the values of the Q factors, where the lowest asymmetry is associated with the maximum values of both parameters.

Symmetry-broken ring structures with varying geometry were fabricated from a 120 nm thick amorphous silicon layer on a glass substrate using electron beam lithography and reactive ion etching (see Methods for details). All measurements shown in the manuscript are based on ring structures with 24 annular unit cells and a base rod length $L_0$ = 335 nm with width $w$ = 115 nm. Symmetry-breaking is introduced by shortening one rod within the unit cell as indicated by the corresponding asymmetry parameter $\Delta L$ (Fig. 2b). The choice of the asymmetries is based on tradeoff between highest Q factors (and associated strong near fields) and the resulting modulation of the radial BIC resonances in the optical experiments. Although higher field enhancements and ultrahigh Q factors above 8000 are numerically predicted for asymmetries approaching $\Delta L$ = 0 nm (see Fig. S2), an optimum parameter range was found for asymmetry values around $\Delta L$ = 25 nm, taking into account nanofabrication accuracy and the noise characteristics of our spectroscopy setup.

Optical characterization of the structures is carried out with a commercial confocal microscopy setup under collimated white light illumination. We first examine the optical response of the radial BIC structures with different radii and find precise tuning of the resonance wavelength via the ring radius $R$ (Fig. 2c). Importantly, our versatile design provides additional degrees of freedom for tailoring the resonance, such as the number of unit cells constituting the ring and the base length $L_0$ of the all-dielectric rods (Fig. S8).

Consistent with numerical simulations, we observe the highest resonance sharpness for the smallest values of the asymmetry $\Delta L$ with Q factors exceeding 500 (Fig. 2d, for fitting details and full spectra see Fig. S9 and Fig. S10). To our knowledge, this is one of the highest Q factors measured for symmetry-broken qBIC resonances in the visible wavelength range. For all asymmetries, the Q factor increases with decreasing radii (Fig. S3), which we attribute to improved coupling between the resonators. This scaling behavior is consistent with the fact that the coupling vanishes for infinite radii with infinite spacing between the resonators.

Strikingly, the annular arrangement of the resonator elements renders the Q factor mostly invariant under rotations of the incident polarization direction of light[33], maintaining a value above 200 throughout the angular range (Fig. 2e). We attribute the observed spectral polarization invariance to the inherent $C_{24}$ symmetry of our ring design which congruently transforms the ring onto itself for a rotation by one unit cell (15 degrees). Furthermore, we do not observe any polarization angle dependence of the spectral response for smaller rotation angles ($0 \leq \varphi \leq 15°$, see Fig. S7). As a result, we conclude that the spectral response of the radial BIC is fully polarization-invariant for any arbitrary polarization angle. Crucially, this polarization-independent performance enables simplified experimental measurements and greatly eases practical device applications.

**Enhanced biomolecular sensitivity**

We leverage the high-Q resonances and strong surface-confined near fields of the radial BIC platform to demonstrate biomolecular sensing for different ring geometries and asymmetries based on a biotin-streptavidin binding bioassay (Fig. 3a). As an initial step, we assess the refractometric sensing performance by varying the local refractive index around the ring structure (see Methods for details) by means of magnetron sputtering of conformal silicon dioxide ($SiO_2$) thin films with increasing thickness. Pronounced resonance shifts for all asymmetries can be detected dependent on the $SiO_2$ layer thicknesses (Fig. 3b). For the calculation of the bulk refractive index sensitivity $S_B$ (BRIS), the resonance shifts mediated by the thin films are converted by considering both the electric near-field decay length of the resonators and the film thickness as introduced previously[34] (see Methods).

Based on these BRIS values, we find that the sensing figure of merit

$$\text{FOM} = S_B/\text{FWHM}, \qquad (1)$$

where FWHM is the resonance full width at half maximum) is clearly correlated with the asymmetry of the radial BIC structures, with the highest FOM of above 20 per refractive index unit (RIU) observed for the lowest asymmetry (Fig. 3c). Despite the much smaller footprint, the FOM values of our radial BIC geometry for biomolecular sensing are mostly comparable with large array-based approaches relying on symmetry protected metasurfaces[18]. Nevertheless, we believe that the FOM can be substantially boosted by utilizing even smaller asymmetries or by further engineering the inter-rod coupling, e.g., by optimizing the resonator shape. Additional advantages of the radial BIC approach can be realized by a concentric arrangement of several rings for multiplexed operation without increasing the footprint, or by integrating multiple rings with identical resonance wavelengths to enhance light absorption efficiency per area.

For the implementation of a model multistep bioassay, we first functionalize our structures with (3-Aminopropyl) triethoxysilane (APTES), followed by the attachment of biotin molecules and the binding of different concentrations of streptavidin protein. After each step, we observe a clear redshift of the radial BIC resonance (Fig. 3d), induced by the higher number of molecules on the surface and the associated increase of the environmental refractive index. To compare the biomolecular sensing performance for all asymmetries, we normalize the resonance shift by the respective FWHM and plot it for each streptavidin concentration to obtain a map of biomolecular sensing performance (Fig. 3e).

Notably, a clear and streptavidin-dependent sensor response over a broad range of concentrations and resonance linewidths is observed. Depending on the sensitivity and spectral resolution of the spectroscopic equipment in experiments and thus the required resonance modulations as well as Q factors, we demonstrate a wide parameter space for picking the appropriate structural asymmetry. Especially, field-deployed applications with low spectral resolution may require lower Q resonances – whereas when pushing for highest sensitivity, high-Q resonances and consequently a high FOM is favored. Our versatile radial BIC platform allows to cover all such use cases in an ultrasmall footprint, pushing the limits of on-chip multiplexing applications for biosensing and opening up new avenues for the precise on-demand tailoring of FOM, Q factor, resonance position, and resonance modulation.

**Enhanced SHG in an evanescently coupled monolayer of MoSe$_2$**

To further demonstrate the versatility of our design, we employed the radial BIC platform for enhanced light-matter interaction between the all-dielectric resonator system and a two-dimensional excitonic material[20,35,36]. Specifically, we demonstrate and localize enhanced second harmonic generation (SHG) in a non-centrosymmetric monolayer crystal of the transition metal dichalcogenide (TMD) MoSe$_2$, facilitated by the highly enhanced electromagnetic near fields that are associated with the radial BIC resonances. For this purpose, we deterministically transferred a monolayer of MoSe$_2$ on top of the radial BIC platform (Fig. 4a). Crucially, this experiment is enabled by the low footprint of the radial BIC platform, which allows full spatial overlap between the micron-sized MoSe$_2$ layer and the ultracompact ring geometry. Atomic force microscopy (AFM) imaging confirms the homogeneous coverage of the resonators (Fig. 4b) and indicates that the monolayer is in good contact with the silicon structures. In Figure 4c, we show the quadratic dependence of the integrated SHG signal on the input power for the monolayer together with its polarization resolved signal (inset, see Methods), confirming the origin of the signal from the non-linear harmonic light generation in the atomically thin crystal[37].

To demonstrate spectrally selective SHG, we excite the coupled system of a MoSe$_2$ monolayer on top of a tailored radial BIC structure at three different excitation wavelengths with the use of a tunable femtosecond pulsed laser (Fig. 4d-e, see Methods for details). For this purpose, we choose a ring with $\Delta L$ = 50 nm to balance the high near-field enhancements with a sufficient resonance modulation. Although we expect even higher near-field and thus SHG enhancements for smaller asymmetries (see Fig. S6), these resonances are often strongly damped by the TMD monolayer in experiments. For our SHG measurements, we excite the system at 744 nm in resonance with the radial BIC mode, at 800 nm close to the MoSe$_2$ exciton peak, and at 900 nm, spectrally distinct from both the exciton and the radial BIC resonance (Figure 4e). At each excitation wavelength, we raster the focused laser beam across the sample, direct the light to an avalanche single photon detector, and plot the obtained SHG signal intensity for each point. In Figure 4d, we show the maps of the SHG signal on the ring normalized to the bare MoSe$_2$ monolayer flake to guarantee consistency between the measurements and to eliminate wavelength-dependent processes such as detector and SHG efficiency as well as transmission of the signal in the optical beam path (see Fig. S11b for raw data without background).

For radial BIC-resonant excitation, we observe a five-fold SHG intensity enhancement confined to the ring structure, illustrating the highly localized generation of the second harmonic signal. In contrast, for off-resonant excitation, we observe a suppression of the SHG (Fig. 4d) within the same region. The suppression of the SHG signal when excited spectrally apart from the radial BIC resonance can be attributed to strain introduced in the monolayer TMD by the ring structure[38], which is outperformed by the highly enhanced near fields of the resonant excitation. The wavelength dependence is a clear indication that the SHG enhancement is driven by the locally enhanced electromagnetic near fields of the radial BIC resonances. We further examined the SHG signal originating from the bare silicon radial BIC structures and observed only a negligible contribution to the overall SHG intensity (see Fig. S11a).

We further conducted control experiments where we suppressed the radial BIC resonance by investigating a symmetric structure ($\Delta L$ = 0 nm, Fig. S12). In this case, we observe no enhancement of the SHG signal when the monolayer is on top of symmetric rings and probed at the wavelength associated with the radial BIC resonance at 744 nm, further confirming that the observed SHG enhancement arises from the near-field coupling of the MoSe$_2$ monolayer with the radial BIC mode. We also perform linearly polarized excitation and show the corresponding polarization dependent SHG maps in Fig. S13. Crucially, the SHG generation experiments are highly facilitated by the compact footprint and polarization invariance of our ring geometry since the constraints on the spatial alignment and orientation of the TMD monolayer are released.

**Discussion**

We have introduced a new platform for enhancing light-matter interactions that combines all benefits of symmetry-broken qBIC approaches while providing at least 50 times less footprint compared to extended 2D metasurfaces[19]. We have demonstrated the enormous potential of our nanophotonic framework by showing highly sensitive biomolecular detection experiments as well as localized SHG enhancement in a monolayer of $MoSe_2$. Our results enable on-demand tailoring of high-Q resonances in ultracompact optical devices for a variety of applications such as integrated multi-resonant and hybrid sensors as well as photonic modes for coupled polariton physics.

**METHODS**

**Simulations.** Numerical calculations were performed in the frequency domain by obtaining full-wave solutions of Maxwell's equations using the commercially available RF module of the finite element solver COMSOL Multiphysics. Port boundary conditions were employed for the excitation of the structures and perfectly matched layers (PML) domains for the absorption of propagating waves. A dispersionless refractive index of $n = 1.45$ was employed for the $SiO_2$ substrate, while the dispersive refractive index of the amorphous silicon film, as measured via ellipsometry, was used for the resonators. The field normalization was performed considering the solution obtained for the sum of the in-plane components of the electric field ($E_x + E_y$) using as a reference the smallest chosen asymmetry ($\Delta L = 25nm$) to highlight the reduction in field enhancement for larger asymmetries (for near field intensities see Fig. S6).

**Radial quasi-BIC fabrication.** Prior to fabrication, the fused silica substrates are cleaned in an ultrasonic bath at 55°C with acetone and rinsed with isopropanol (IPA) afterwards followed by an oxygen plasma etching to guarantee cleanliness of the substrates. Afterwards, a 120 nm thick amorphous silicon (a-Si) is deposited at 250° C via plasma-enhanced chemical vapor deposition on top of the silica substrate. We used a 180 nm thick double layer of the positive tone resist poly (methyl methacrylate) (PMMA) with different chain lengths (495k and 950k) to guarantee an undercut of the resist after development which aids for the liftoff process. We baked each resist layer at 170°C for 3 minutes. A conductive polymer (E-Spacer 300Z) is deposited on top of the PMMA double layer to suppress charge accumulations on the substrate.

Electron beam lithography (EBL) is utilized to define the nanostructure pattern into the resist using 30 kV acceleration voltage for maximum edge sharpness of our nanostructures. The illuminated resist is developed in a 3:1 IPA: MIBK (Methyl isobutyl ketone) for 90 seconds after the organic polymer has been washed of by a water bath for 10 seconds. We use a combination of 20 nm silicon dioxide ($SiO_2$) and 25 nm chromium (Cr) layer deposited via electron beam deposition as hard mask where the $SiO_2$ layer serves as an effective diffusion barrier for the Cr atoms which can introduce severe losses within the silicon radial BIC structures. The liftoff of the metal structures is performed in special remover (Microsposit remover 1165) at 80°C overnight. The hard mask pattern is transferred into the silicon film via inductively coupled reactive ion etching (ICP-RIE) by chlorine and fluorine chemistry. After removing the Cr hard mask with a chromium wet etch, we remove the $SiO_2$ hard mask by a final ICP-RIE step to obtain the pure silicon nanostructures.

**Optical measurements.** The optical characterization of our radial BIC structures is conducted with a commercial white light transmission microscopy setup (WiTec alpha300 series). Importantly, the samples are illuminated with collimated white light from the backside and the transmitted signal is collected with a 50x objective (NA = 0.8). Because of the collimated and linearly polarized illumination, good agreement can be obtained with the simulations, where the same illumination conditions are assumed. The collected light is focused into a fiber which is coupled to a grating-spectrometer equipped with a silicon CCD camera that we use to acquire white light transmittance spectra. All the transmittance spectra shown are referenced to the bare $SiO_2$ substrate. We note that the interaction cross-section of the white light depends on the ring radius of the constituent geometry, thus, different radii lead to different transmitted white light intensities potentially causing slightly different resonance modulations. This behavior, however, does not impede our physical observations and our conclusions are in excellent agreement with the theory.

**Biofunctionalization.** The radial BIC structures were treated by piranha solution and oxygen plasma cleaning processes to activate the structure surface with a high amount of hydroxyl radicals (-OH). Then a silanization protocol using (3-aminopropyl)triethoxysilane (APTES, 10% v/v in ethanol, overnight) was used to modify the structure surface with $NH_2$ groups, which is ready to capture sulfo-NHS-biotin molecules (ThermoFisher Scientific, Mn=443.4, as a linker between the radial resonators and target molecules). The sulfo-NHS-biotin molecules dissolved in phosphate buffered saline (1×, PBS) buffer solution were drop-casted on the structures with a volume of 40 μL and allowed to incubate for 2 hours at room temperature, followed by a rinsing step using 1× PBS buffer and drying under $N_2$ flow. Afterwards, the biotin-immobilized resonators were incubated in target analyte solutions of streptavidin with varying concentrations for 2 hours at room temperature. Once the streptavidin incubation finished, the resonators were rinsed with PBS buffer, dried using $N_2$ flow, and transferred to the optical characterization setup.

**Determination of quality factors.** For the Q factor extraction of our sharp radial BIC resonances, we employ temporal coupled mode theory[39] to extract the linewidth $\gamma$ of the resonance (see Fig. S9) using

$$T = \left| i\, e^{i\phi} t_0 + \frac{\gamma_r}{\gamma_r + \gamma_i + i(\lambda - \lambda_{\text{res}})} \right|^2 \qquad (2)$$

where $e^{i\phi}t_0$ accounts for the background transmission and controls the shape of the resonance. The Q factor is calculated as[32]

$$Q = \frac{\lambda_{\text{res}}}{2\gamma}, \qquad \gamma = \gamma_i + \gamma_r, \qquad (3), (4)$$

where $\lambda_{\text{res}}$ is the resonance wavelength and $\gamma_i$ and $\gamma_r$ is the intrinsic and radiative loss rate, respectively. The same fitting approach is applied for the simulated and experimental transmittance spectra to ensure consistent Q factors.

**Extraction of bulk refractive index sensitivity.** The bulk refractive index sensitivity is extracted from the surface sensitivity $S_S$, which can be calculated using

$$S_S = \frac{\Delta\lambda_{\text{res}}}{\Delta n_S}, \qquad (5)$$

where $n_S$ is the refractive index of the conformally deposited $SiO_2$ thin film[34]. Similar to the bulk figure of merit for refractive index sensing, we define the figure of merit for surface sensitivity as

$$\text{FOM}_S = \frac{S_S}{\text{FWHM}}, \qquad (6)$$

with FWHM being the full width at half maximum of the radial BIC resonance. From the figure of merit for surface sensitivity, we can calculate the figure of merit for bulk refractive index sensitivity via

$$\text{FOM}_B = \frac{\text{FOM}_S}{1 - \exp(-2h/l_d)}, \qquad (7)$$

where $h$ is the thickness of the thin films and $l_d$ is the decay length of the electric near fields that we extract from simulations as 44.7 nm.

**2D material fabrication and transfer**. The exfoliation of large area atomically thin monolayers of molybdenum diselenide ($MoSe_2$) is done via mechanical exfoliation from a seed crystal (HQ Graphene) with blue tape (Nitto) onto polydimethylsiloxane (PDMS) stamps. Suited monolayers for transfer are identified with an optical microscope and monolayers are confirmed by photoluminescence imaging on the PDMS substrate. For the transfer, we use a home-built transfer setup with motorized z stage to control the stamp movement.

**SHG measurements.** We employed the output of a tunable Ti: sapphire laser (Coherent Chameleon Ultra II) to selectively excite the SHG signal from the $MoSe_2$ monolayer sample mounted in an inverted microscope setup and collected in a back reflection geometry. The sample is placed on a motorized stage allowing the mapping of the sample surface. The SHG signal is isolated from the excitation beam with a short pass filter (Semrock) and collected with an avalanche photodiode detector (Micro Photon Devices). The polarization resolved SHG signal is obtained by rotating the linear polarization of the excitation beam with a halfwave plate. The signal is then filtered with a linear polarizer, either parallel or perpendicular to the initial excitation laser polarization, before being collected in a grating-spectrometer equipped with a silicon CCD camera (Princeton Instruments).

**Data availability**

The main data supporting the findings of this study are available within the article and its Supplementary Information files. Extra data are available from the corresponding author upon reasonable request.

## Acknowledgements

The authors thank Philipp Altpeter for fabrication advices, Thomas Weber for the implementation of the fitting routines, as well as Christoph Hohmann for support with rendered images. Our studies were funded by the Deutsche Forschungsgemeinschaft (DFG, German Research Foundation) under grant numbers EXC 2089/1 – 390776260 (Germany´s Excellence Strategy) and TI 1063/1 (Emmy Noether Program), the Bavarian program Solar Energies Go Hybrid (SolTech), the Center for NanoScience (CeNS), the Australian Research Council (the grant DP210101292), the US Army International Office (grant FA5209-21-P0034), and the National Council for Scientific and Technological Development (CNPq)(PDJ 2019 – 150393/2020-2). S.A.M. additionally acknowledges the EPSRC (EP/W017075/1) and the Lee-Lucas Chair in Physics. L. S. and H. R. acknowledge funding support from the DECRA Project (DE220101085) from the Australian Research Council.

## Author contributions

L. K. and A. T. conceived the idea and planned the research. L. K. fabricated the radial BIC samples. L. K., L. S., J. W., and H. R. performed optical measurements. L. K. and A. T. analyzed the data. R. B. did numerical simulations. R. B. and Y. K. performed the theoretical analysis of the radial BIC platform. S. M., Y. K., and A. T. supervised the project. All authors discussed the results and contributed to the writing of the manuscript.

## Competing interests

The authors declare no competing interests.

**Figures**

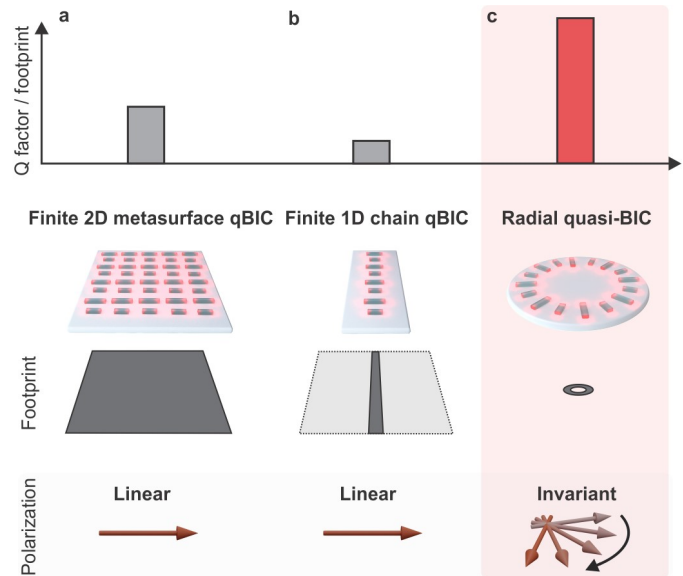

**Fig. 1 | Conceptual advantages of radial quasi-bound states in the continuum (radial BICs).** Established symmetry-broken quasi-BIC geometries, such as 2D metasurfaces (**a**), and 1D chains (**b**), exhibit large footprints, moderately high Q factors and require polarization-dependent excitation. The radial BIC platform (**c**) combines a tiny footprint with high Q factors in the visible. Above all, the radial BIC platform provides the highest Q factor per footprint ratio compared to other 1D and 2D BIC-based platforms as shown here.

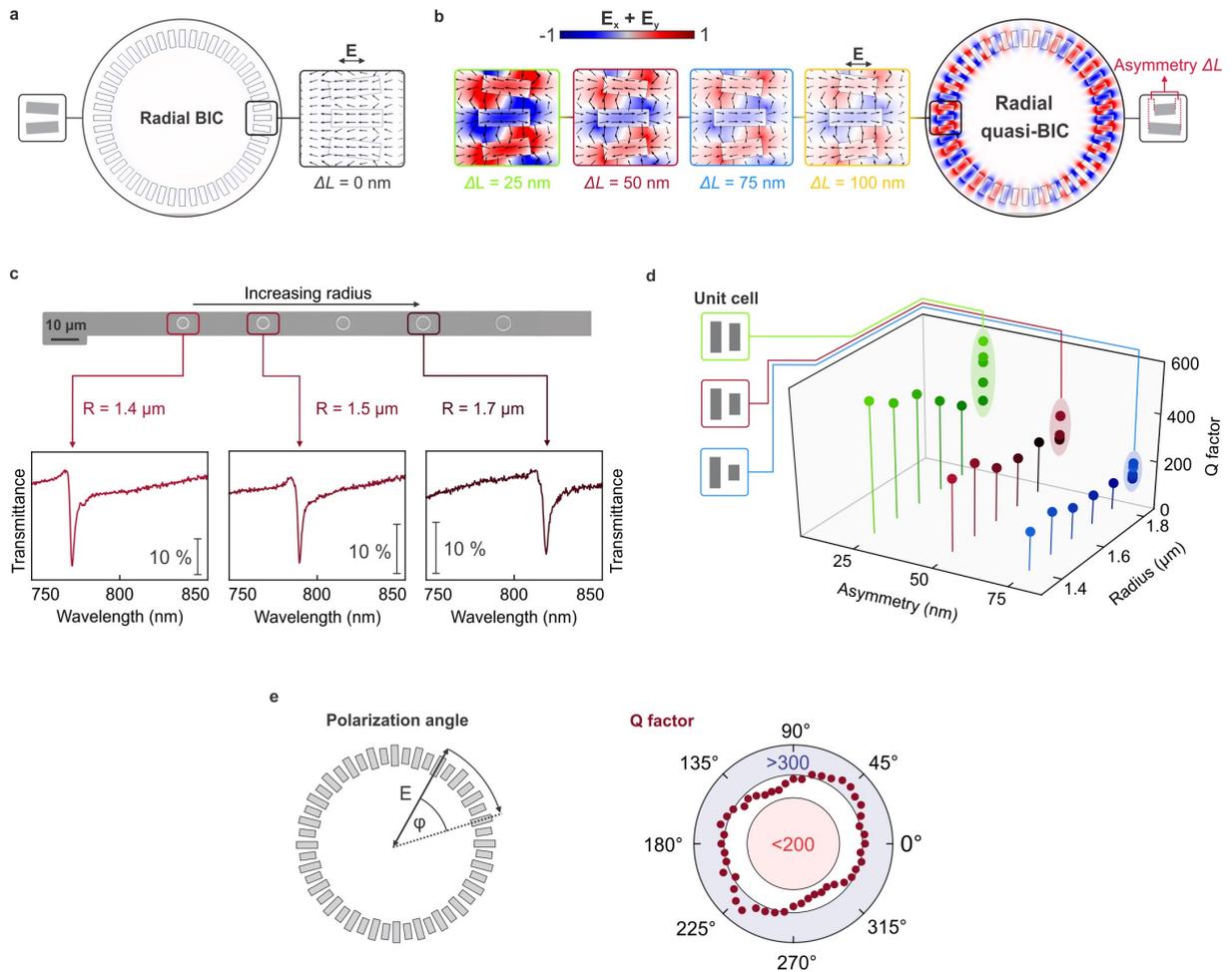

**Fig. 2 | Versatile radial BIC resonances and polarization invariance.** Electric near fields for a radial BIC ($\Delta L = 0$ nm) in (**a**) and several symmetry-broken radial quasi-BIC ($\Delta L > 0$ nm) geometries in (**b**). **c**, Optical transmittance spectra of the radial BIC structures show a resonance redshift with increasing radii as illustrated in the grey-scale optical micrograph. **d**, Dependence of the quality factor on the ring radius $R$ and the unit cell asymmetry $\Delta L$. Quality factors exceed 500 for $\Delta L = 25$ nm in the visible wavelength range. **e**, The radial BIC geometry shows polarization invariance as observed by the weak dependence of the resonance quality factor on the polarization angle of the incident light $\varphi$.

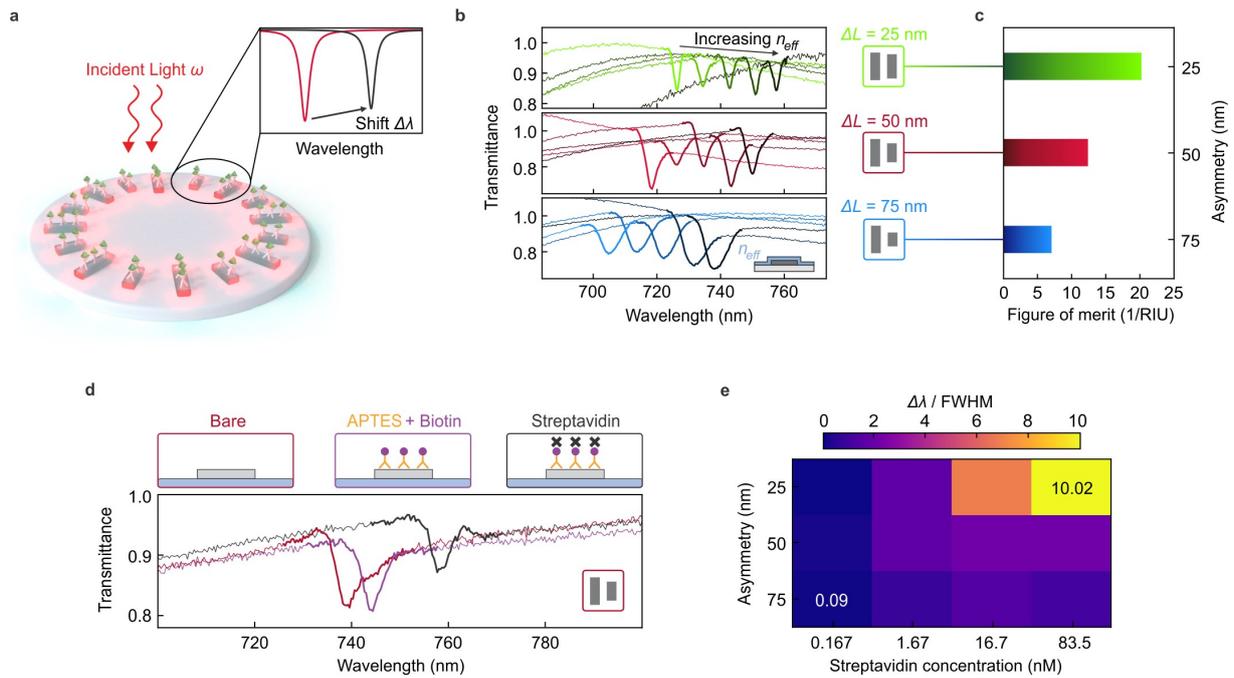

**Fig. 3 | Refractive index and molecular biosensing with radial BICs. a,** Sketch of the biosensing experiments. The ring structures are functionalized with capture antibodies and shifts in the spectral position of the resonance are recorded for the binding of different concentrations of biomolecules. **b,** Optical transmittance spectra of three-unit cell asymmetries $\Delta L$ with $R = 1.5\ \mu m$ covered with different thicknesses of conformal SiO$_2$ thin films. **c,** Corresponding figure of merit for bulk refractive index sensing. **d,** Transmittance spectra for a ring with $\Delta L = 50$ nm and $R = 1.6\ \mu m$ after each functionalization and molecular binding step as indicated in the color-coded boxes. **e,** Map of biomolecular sensing performance. Measured resonance shifts normalized to the respective FWHM for three different asymmetries dependent on streptavidin concentrations.

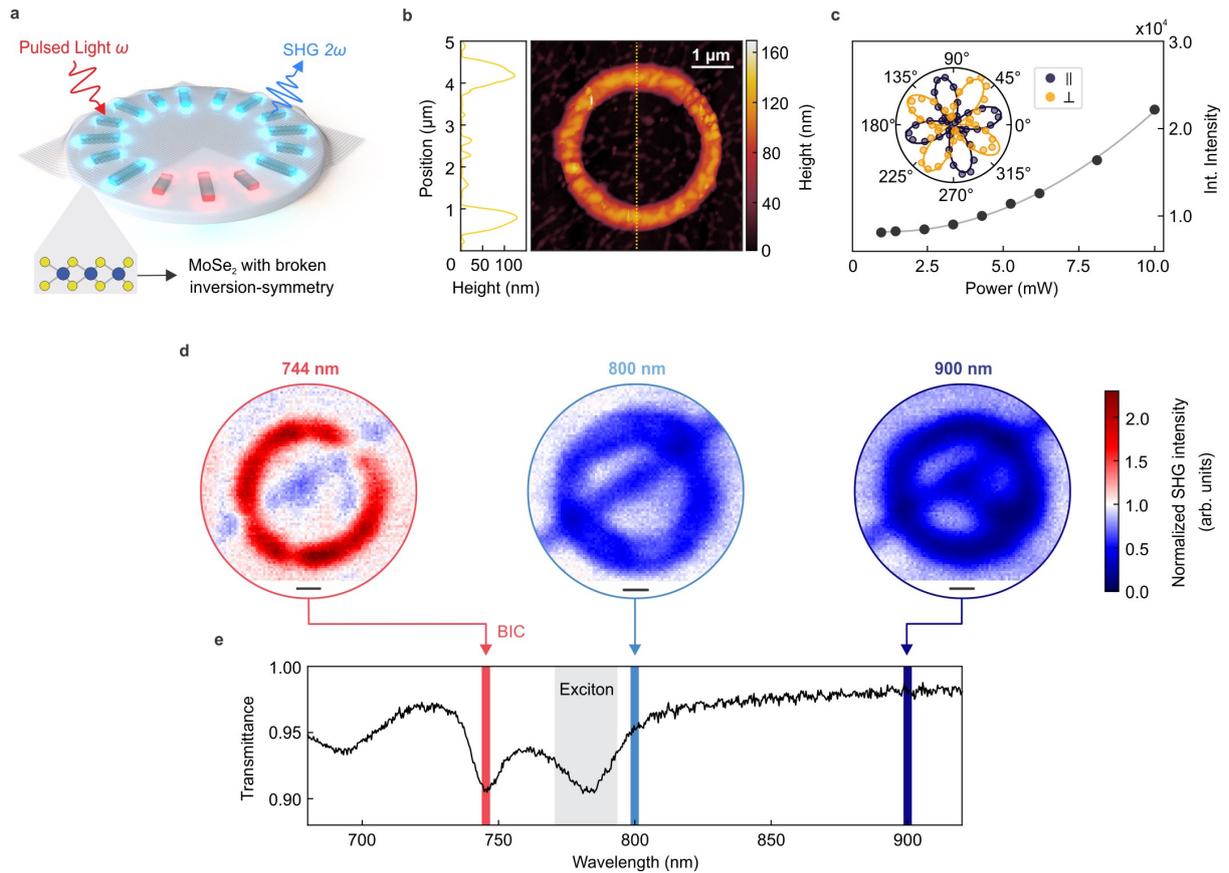

**Fig. 4 | Radial BIC-enhanced second-harmonic generation in a MoSe₂ monolayer. a,** Schematic depiction of a MoSe₂ monolayer covering the ring structures while being illuminated by a pulsed excitation laser. **b,** AFM image showing the coverage of the flake on top of the ring. **c,** Integrated second harmonic generation (SHG) signal for an excitation wavelength of $\lambda$ = 744 nm on the MoSe₂ monolayer on a bare substrate showing a quadratic dependence (grey fit line). Inset: Polarization-resolved SHG signal from the MoSe₂ monolayer in either parallel or perpendicular detection (see Methods). **d,** SHG maps for the ring displayed in panel **b** taken at different excitation wavelengths. SHG enhancement is only present for an excitation wavelength resonant with the radial BIC. In contrast, for the off-BIC excitation, we observe the expected suppression of the SHG signal due to strain. Scale bar: 300 nm. **e,** Transmittance spectrum of the ring structure ($\Delta L$ = 50 nm) covered with MoSe₂ monolayer clearly showing the spectral signatures of the radial BIC resonance (744 nm) next to the absorption line of the exciton (785 nm).

**Supplementary Information**

# Radial bound states in the continuum for polarization-invariant nanophotonics

**Lucca Kühner[1,2], Luca Sortino[1,2], Rodrigo Berté[1,2,3], Juan Wang[1,2], Haoran Ren[4],**

**Stefan A. Maier[4,5,1,2], Yuri Kivshar[6] and Andreas Tittl[1,2,\*]**

[1]Chair in Hybrid Nanosystems, Nanoinstitute Munich, Faculty of Physics, Ludwig-Maximilians-Universität München, Königinstrasse 10, 80539 Munich, Germany

[2]Center for NanoScience (CeNS), Faculty of Physics, Ludwig-Maximilians-Universität München, Schellingstrasse 4, 80799 Munich, Germany

[3]Instituto de Física, Universidade Federal de Goiás, 74001-970 Goiânia-GO, Brazil

[4]School of Physics and Astronomy, Monash University, Clayton, Victoria 3800, Australia

[5]The Blackett Laboratory, Department of Physics, Imperial College London, London, SW7 2AZ, United Kingdom

[6]Nonlinear Physics Centre, Research School of Physics Australian National University, Canberra ACT 2601, Australia

*e-mail: andreas.tittl@physik.uni-muenchen.de

**Supplementary note 1: Comparison of quality factors in different qBIC systems**

There are numerous approaches for realizing quasi bound states in the continuum (qBICs) in photonic metasurfaces, which are all accompanied by certain benefits and drawbacks. We distinguish three classes here for simplicity: accidental and photonic-crystal-based qBICs, qBICs emerging from strongly coupled photonic modes in individual subwavelength resonators, and symmetry-broken qBICs. So far, photonic-crystal-based approaches have shown the highest quality factors, with specific examples reaching around $10^6$ [1]. However, they often rely on grating-based approaches and lack flexible tunability of the resonances. Quasi-BICs emerging from strongly coupled modes possess the lowest footprints since the do not rely on array-based approaches [2]. Nevertheless, these resonators have so far been limited to experimental Q factors below 200, often require complex optical excitation with structured light, and exhibit near-fields that are mostly located inside the structures, reducing the spatial overlap with analytes and thus the potential sensing performance.

In contrast, symmetry-broken BICs show exceptional resonance tunability and sustain high-Q resonances down to small array sizes, with record Q-factor values of roughly 18500 [3]. For better comparison, we show some representative works in the field of BIC-based nanophotonics in Table 1. Notably, the radial BIC exhibits the smallest footprint and thus highest Q factor vs. footprint ratio apart from the single disk BIC, which we mentioned above but exclude here due to the complex optical excitation conditions and unsuitable near-field distributions for sensing applications. [4,5] In fact, comparing the same number of unit cells with Ref. [3], the radial BIC exhibits the same quality factor although confined to a much smaller footprint and sustains resonances in the visible where material losses and fabrication imperfections – in contrast to the NIR region – pose significant challenges. In fact, we carried out simulations for our radial BIC geometry, which demonstrate that the ring arrangement of unit cells is favored compared to the arrangement in a 2D array (see Figure S3). We attribute this effect to the improved inter-resonator coupling.

| | Ref. 5 | Ref. 3 | Ref. 4 | Ref. 1 | Ref. 2 | Our work |
|---|---|---|---|---|---|---|
| BIC mechanism | Symmetry-broken | Symmetry-broken | Accidental | Photonic crystal | Strong mode coupling | Symmetry-broken |
| Excitation | Linear pol. | Lin. Pol. | Lin. Pol. | Lin. Pol | Structured | Lin. Pol. |
| Resonance wavelength | **VIS**, 750-860 nm | **NIR**, 1490 – 1550 nm | **VIS/NIR**, 825 – 842 nm | **VIS**, 570-850 nm | **IR**, 1500-1600 nm | **VIS**, 700-850 nm |
| Quality (Q) factor | 144 | 500 - 18511 | 2750 | $10^6$ | Up to 200 | 100 - 500 |
| Footprint | 40000 µm² | 10.6 – 308* µm² | 24000 µm² | $10^8$ µm² * | 0.62 µm² | 2.7 µm² |
| Q / footprint | 0.0036 | 66 - 117 | 0.11 | 0.01 | 322 | 186 |

*Marked fields were not explicitly given in the manuscript and were estimated from images or spectra.

**Supplementary note 2: Tradeoff between electric near-field and resonance modulation**

For a radial BIC geometry composed of a lossless dielectric, we expect higher near-fields for decreasing asymmetries, with a maximum value for an infinitesimal deviation from the symmetric case (see, e.g., Ref. [7]), and thus the near-field intensity should peak for the smallest possible value of $\Delta L$, which is still larger than zero. To illustrate this behavior, we added a plot showing the associated near fields for different rings down to $\Delta L$ = 5nm in Figure S5.

As is obvious from the plots, the near fields increase tremendously for reduced asymmetries, but enter a saturation regime when going below an asymmetry of $\Delta L$ = 25 nm. As a result, a further reduction of the asymmetry will not be accompanied by significant increase of the near fields and thus, the chosen asymmetry is already in the optimum region.

At the same time, in real experiments, there is competition between radiative and parasitic losses, such as material losses, losses induced by surface or edge roughness or even losses related to the statistical variation of the geometrical cross sections of the constituent resonators within the unit cell [8]. These parasitic losses fundamentally limit the achievable quality factors but also the maximum achievable resonance modulations and will consequently diminish resonance modulations for asymmetries below 25 nm.

Our reason to choose a 25 nm asymmetry in experiments was to balance high near field enhancements with sufficient resonance modulation at the experimental condition nonzero losses such that the resonance could still be retrieved reliably at our white light confocal microscope. Taken together, the minimum asymmetry which we used was the optimum value for a finite but not infinitesimally small value of $\Delta L$, see a similar case explained in the connection with for the nonlinear effects (see Ref. [9]).

**Supplementary figures**

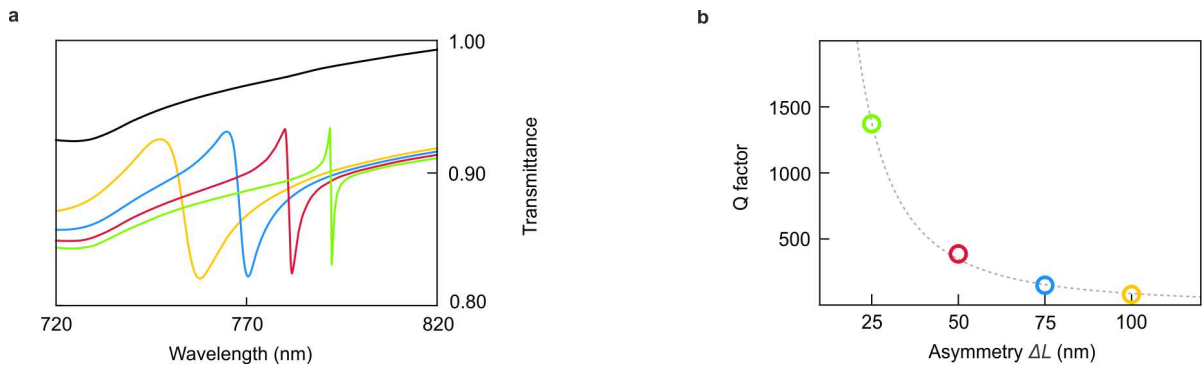

**Fig. S1 | Simulated spectral radial BIC responses for different asymmetries. a,** Simulated transmittance spectra of radial BIC structures with different asymmetries with $R$ = 1.5 µm, where the black response (shifted for clarity) corresponds to the symmetric case. **b,** Extracted Q factors for the spectra shown in (**a**) follow the well-known $(\Delta L)^{-2}$ behavior (fitted as gray dashed line).

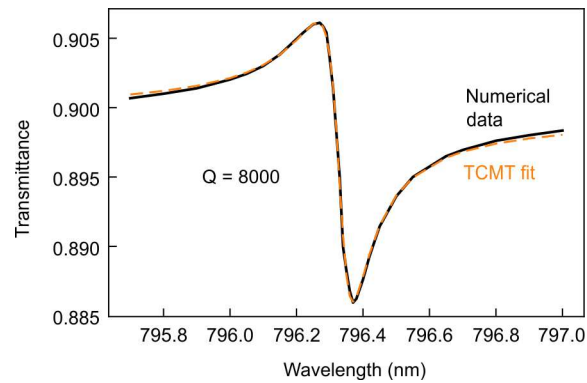

**Fig. S2 | Towards the highest Q factors of the radial BIC platform.** Numerical data for an asymmetry of $\Delta L$ = 5 nm and a ring radius of $R$ = 1.5 µm show Q factors above 8000.

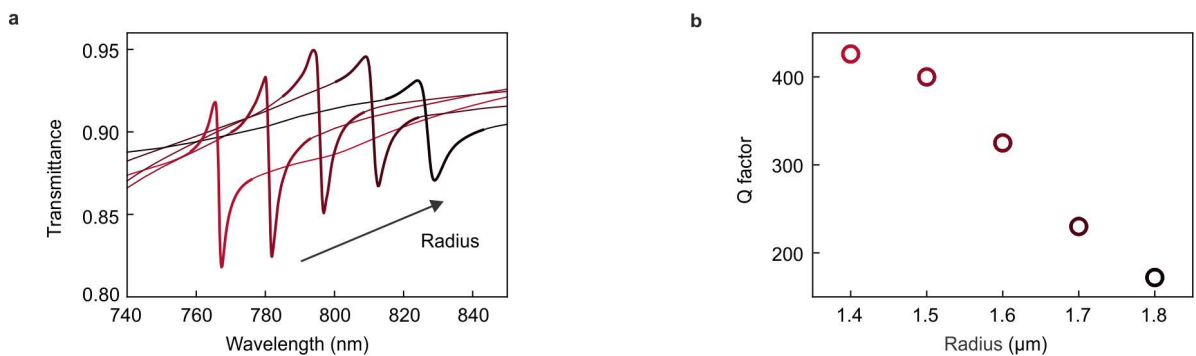

**Fig. S3 | Simulated spectral radial BIC responses for different radii. a,** Simulated transmittance spectra of radial BIC structures with different radii ranging from $R$ = 1.4 µm to $R$ = 1.8 µm for $\Delta L$ = 50 nm. **b,** Extracted Q factors for the spectra shown in (**a**), showing the highest Q factor for the smallest radius.

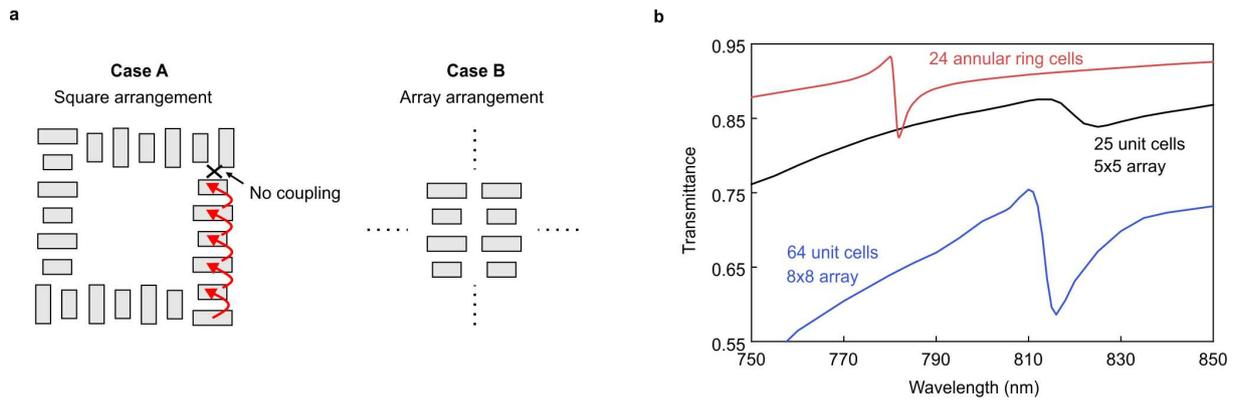

**Fig. S4 | Advantages of an annular arrangement of unit cells. a,** Different possible arrangements for polarization invariant response (case A) that lacks mode coupling at the edges of the structure and reduced footprints (case B) without polarization invariance and worse performance (see panel b). **b,** Numerical investigation of the difference between resonator arrays of 8x8 (blue curve) and 5x5 (black curve) unit cells arranged in a two-dimensional array compared to 24 unit cells (red curve) arranged in a ring fashion. Clearly, the resonance is sharpest for the ring arrangement.

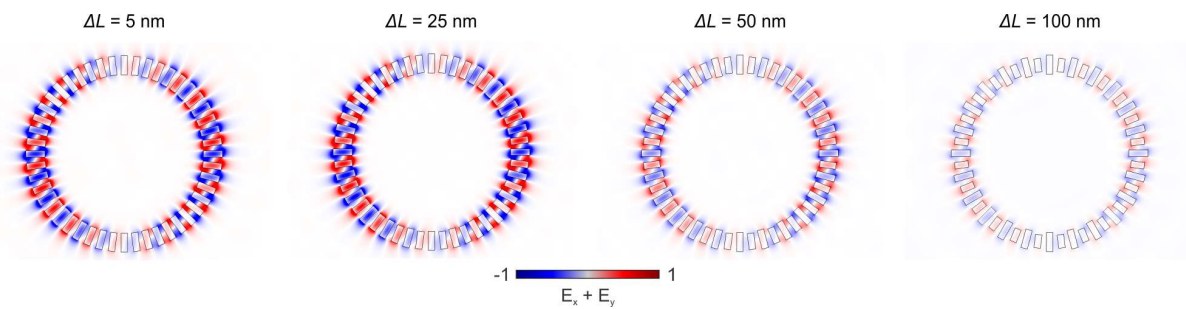

**Fig. S5 | Electric near fields for different ring asymmetries.** Numerical simulations of the electrical near fields for rings starting with $\Delta L$ = 100 nm (right hand side) down to $\Delta L$ = 5 nm. The electric near-fields almost saturate below $\Delta L$ = 25 nm showing only slight differences between the smallest asymmetries. All near fields are plotted on the same scale.

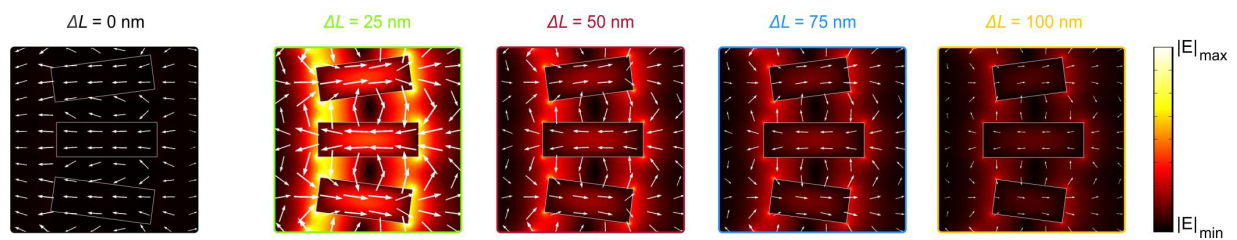

**Fig. S6 | Normalized near-field intensities for different asymmetries.** Numerical simulations of the near-field intensity for the asymmetries given in Fig. 2 in the manuscript. For better comparison, the near-field intensities are all normalized to the same value.

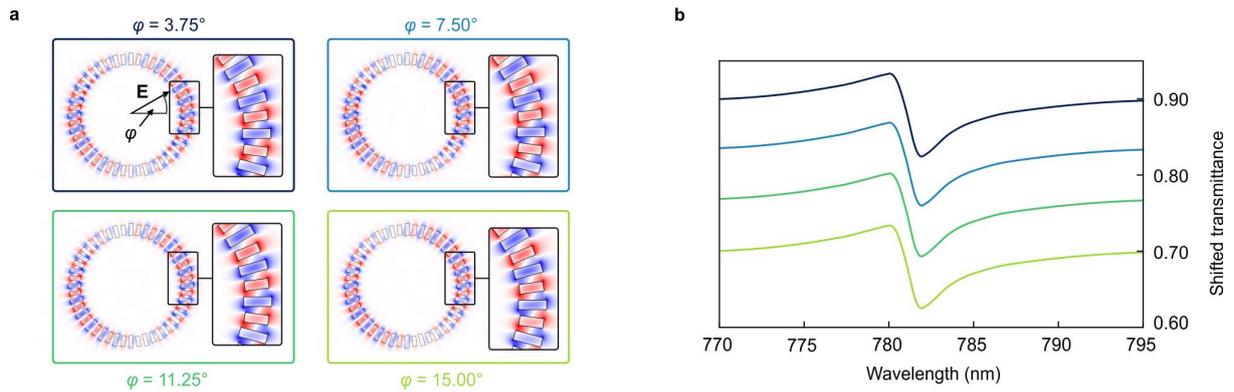

**Fig. S7 | Influence of the polarization angle on radial BIC mode formation. a,** Electric near fields for different linear polarizations $\varphi$ of the incident light showing no differences in the mode formation. **b,** Corresponding numerical spectra for the different polarization angles (shifted for clarity) exhibit the same resonance position, Q factor, and resonance modulation.

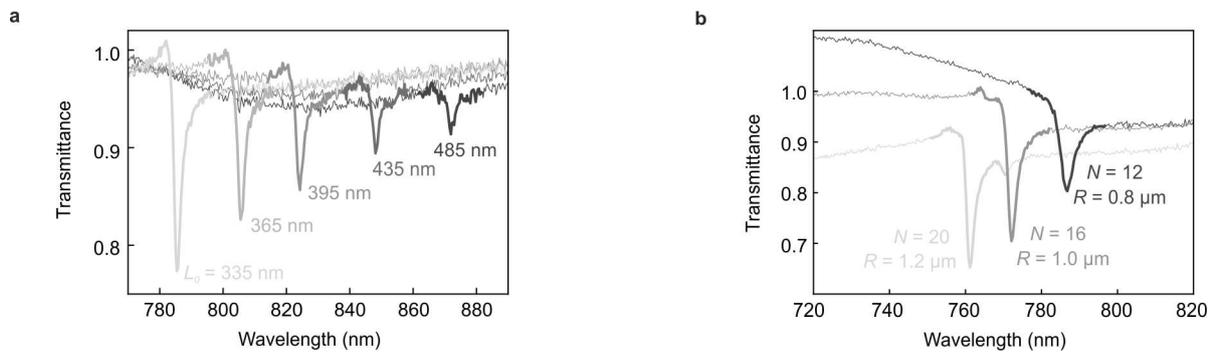

**Fig. S8 | Influence of base rod length $L_0$ and number of unit cells on the optical response of radial BICs. a,** Measured transmittance spectra for increasing base rod length $L_0$ for an asymmetry of $\Delta L$ = 50 nm. A red-shift of the radial BIC resonance is observed for higher base length. **b,** White light transmittance spectra for different number of unit cells with $\Delta L$ = 50 nm. The Q factor decreases for fewer unit cells due to decreased inter-rod coupling.

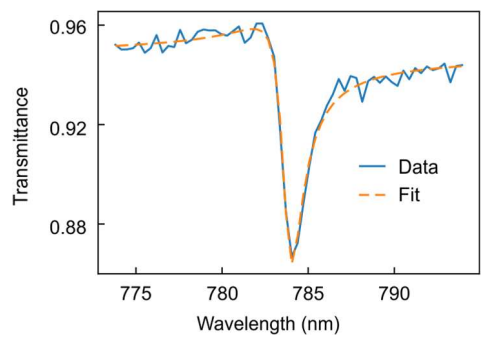

**Fig. S9 | Exemplary fit of measured radial BIC transmittance spectra with the TCMT model.** A high quality factor above 500 in the visible spectral region was determined for a symmetry-broken ring structure with $\Delta L$ = 25 nm and $R$ = 1.4 μm.

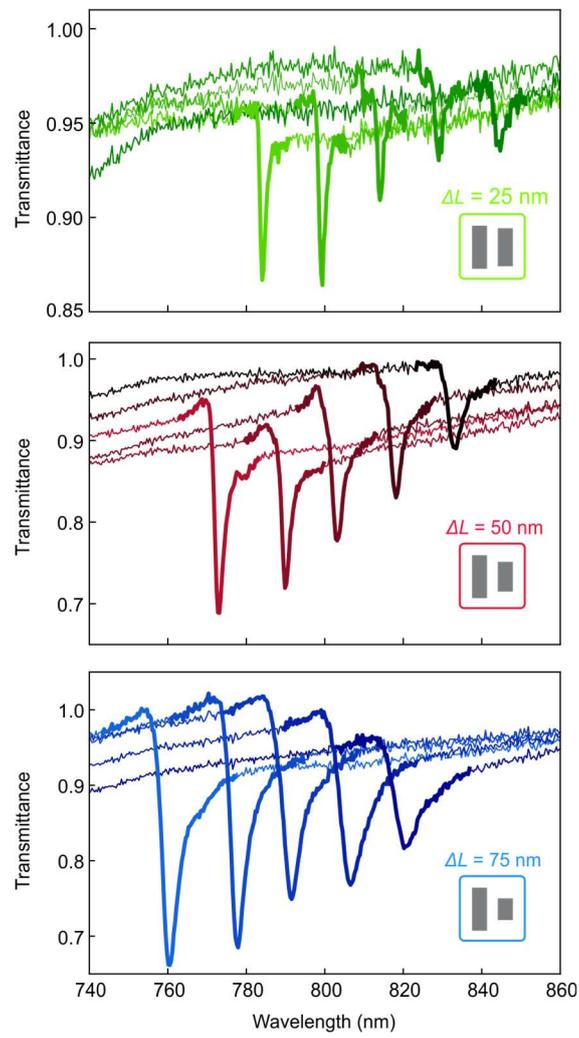

**Fig. S10 | White-light transmittance spectra for different asymmetries.** Optical transmittance spectra of the radial BIC geometry with different asymmetries for a ring radius of $R$ = 1.5 µm. For the smallest asymmetries, we observe the highest Q factors exceeding 500.

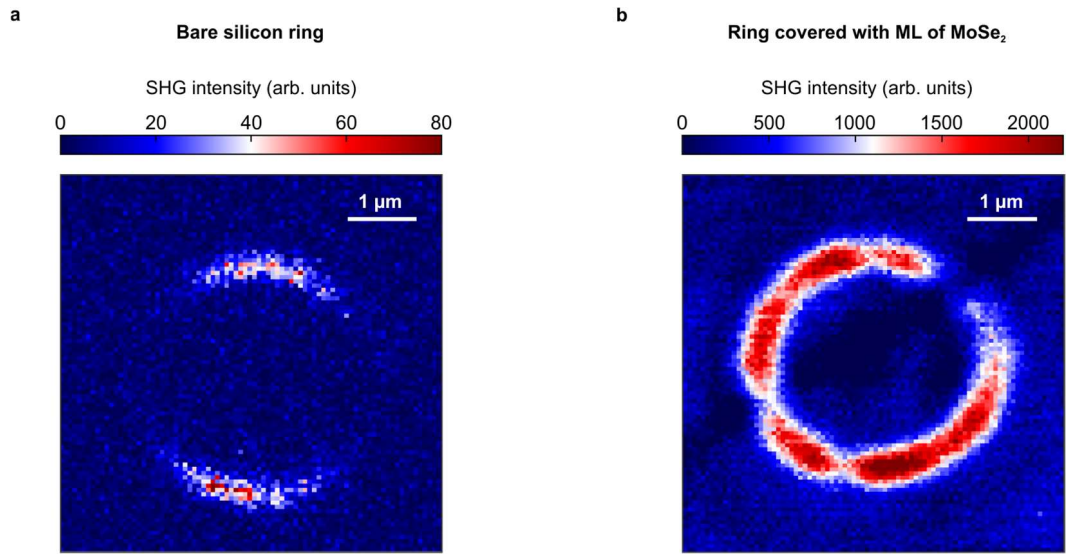

**Fig. S11| SHG map comparison of a bare silicon radial BIC structure and a MoSe₂ covered ring.**
Comparison of SHG maps extracted from a bare asymmetric silicon radial BIC geometry in (**a**) and from the same ring covered with a monolayer of MoSe₂ in (**b**) plotted on different scales. For comparability of both maps, the background signal is set to zero. The maximum signal obtained from bare silicon structures is more than two orders of magnitude smaller on the whole ring compared to the MoSe₂ covered ring and can thus be neglected.

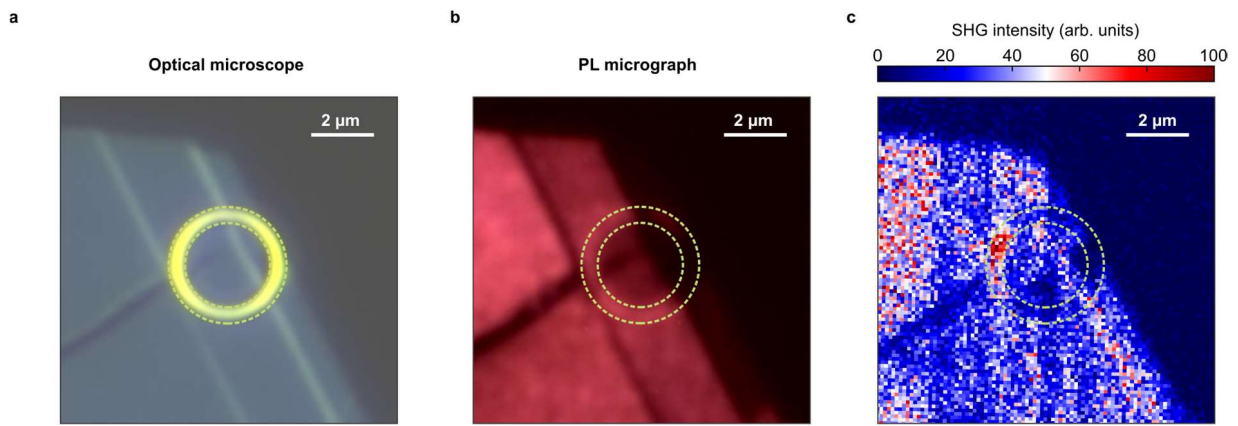

**Fig. S12 | SHG map of a symmetric ring covered with a monolayer of MoSe₂. a, b,** Optical and photoluminescence micrograph of the radial BIC ring which is used for SHG reference maps for the purely symmetric structure. The outline of the ring is indicated as yellow dashed circles for clarity. **c,** SHG map for the symmetric structure. Except for regions where the monolayer is folded, we observe no enhancement of the SHG signal present supporting the BIC-induced SHG enhancement.

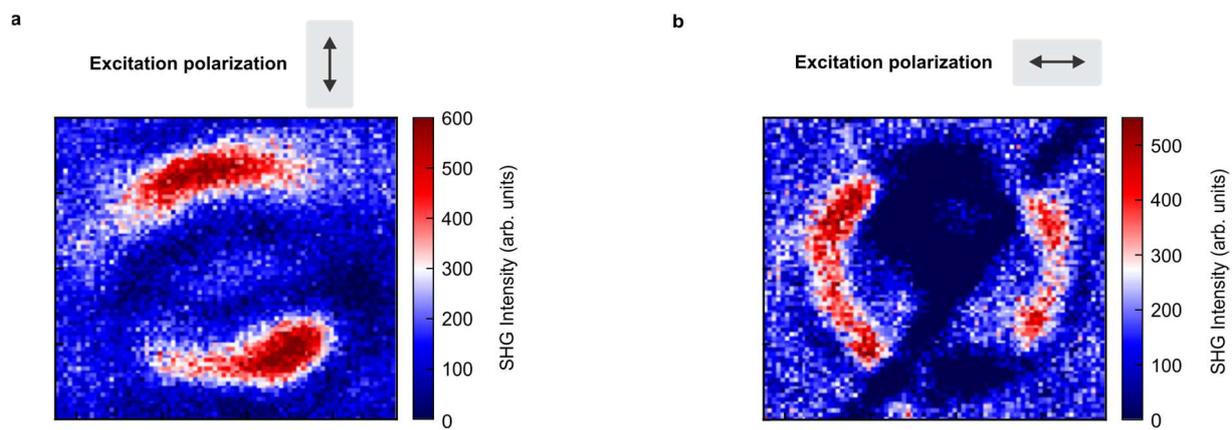

**Fig. S13 | Excitation polarization dependent SHG map of the same ring covered with a monolayer of MoSe₂. a,** Background-corrected SHG map for vertical excitation polarization. **b,** Background-corrected SHG map for horizontal excitation polarization.

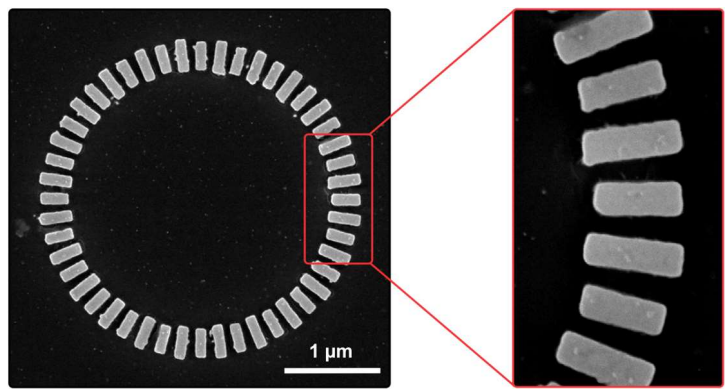

**Fig. S14 | Scanning electron micrographs of the fabricated structures.**